\documentclass[aps,amsmath,amssymb,preprint,groupedaddress,showkeys]{revtex4-2}
\usepackage{tabularx}
\usepackage{graphicx}% Include figure files
\usepackage{dcolumn}% Align table columns on decimal point
\usepackage{bm}% bold math

\usepackage[T1]{fontenc}
\usepackage{multirow, lipsum, booktabs, tabularx}
\usepackage{siunitx, xcolor}
\usepackage{subfigure}
\usepackage[version=4]{mhchem}
\usepackage{color}
\usepackage[colorlinks,linkcolor=blue,anchorcolor=blue,citecolor=blue]{hyperref}

\usepackage{lineno}
\usepackage{color}
\usepackage{soul}
\soulregister\cite7 
\soulregister\ref7 
\sethlcolor{yellow}
\begin{document}

\title{Probing Higgs $CP$ properties at the CEPC}

\author{Qiyu sha} 
\email{shaqiyu@ihep.ac.cn}
\author{Abdualazem Fadol}
\email{amohammed@aims.ac.tz}
\author{Fangyi Guo}
\email{guofangyi@ihep.ac.cn}
\affiliation{Institute of High Energy Physics, 19B, Yuquan Road, Shijing District, Beijing, China, 100049}
\affiliation{University of Chinese Academy of Sciences (CAS)}
\author{Gang Li}
\email{li.gang@ihep.ac.cn}
\affiliation{Institute of High Energy Physics, 19B, Yuquan Road, Shijing District, Beijing, China, 100049}
\affiliation{University of Chinese Academy of Sciences (CAS)}
\author{Jiayin Gu}
\email{jiayin\_gu@fudan.edu.cn}
\affiliation{Department of Physics and Center for Field Theory and Particle Physics, Fudan University, Shanghai 200438, China}
\affiliation{Key Laboratory of Nuclear Physics and Ion-beam Application (MOE), Fudan University, Shanghai 200433, China}
\author{Xinchou Lou}
\email{xinchou@ihep.ac.cn}
\affiliation{Institute of High Energy Physics, 19B, Yuquan Road, Shijing District, Beijing, China, 100049}
\affiliation{University of Chinese Academy of Sciences (CAS)}
\affiliation{University of Texas at Dallas, Richardson, Texas 75080-3021, USA}
\author{Yaquan Fang}
\email{fangyq@ihep.ac.cn}
\affiliation{Institute of High Energy Physics, 19B, Yuquan Road, Shijing District, Beijing, China, 100049}
\affiliation{University of Chinese Academy of Sciences (CAS)}

%\date{\today}

\begin{abstract}
In the Circular Electron Positron Collider (CEPC), a measurement of the Higgs \textit{CP} mixing through $e^{+} e^{-} \rightarrow Z H \rightarrow l^{+} l^{-}(e^{+} e^{-} /\mu^{+} \mu^{-}) H(\rightarrow b \bar{b} / c \bar{c} / g g)$ process is presented, with $5.6\ \mbox{ab}^{-1}$ $e^{+} e^{-}$ collision data at the center-of-mass energy of $240\ \mathrm{GeV}$. In this study, the \textit{CP}-violating parameter $\tilde{c}_{Z \gamma}$ is constrained between the region of $ -0.30$  and $0.27$ and  $\tilde{c}_{Z Z}$  between 
$-0.06$ and $0.06$ at $68\%$ confidence level. This study demonstrates the great potential of probing Higgs $CP$ properties at the CEPC. 
\end{abstract}



\maketitle

\newpage

\clearpage
\section{\label{sec:1}Introduction}
The historic discovery of Higgs boson with a mass around $125\ \mathrm{GeV}$ in 2012 by the ATLAS and CMS collaborations at the Large Hadron Collider (LHC) \cite{ATLAS:2012,CMS:2012} completed the Standard Model (SM).
This particle provides a new portal to search for new physics beyond the SM (BSM). 
The Higgs boson is predicted to be a scalar particle ($J^{P}=0^{++}$) under the SM of particle physics. 
As a result, any observation of charge-parity violation (\textit{CPV}) in Higgs would be a sign of physics BSM and could account for the explanation of the observed baryon asymmetry of the universe.\par

At present, the Higgs boson \textit{CP}-mixing measurements are performed at hadron colliders. 
The hypothesis of spin-1 or spin-2 Higgs has been excluded by ATLAS and CMS at $99 \%$ confidence level (CL) with $\sqrt{s} = 7~\mbox{and}~8$ $\mathrm{TeV}$, $25\ \mbox{fb}^{-1}$ data \cite{ATLAS:2015zhl}.
Studies of the \textit{CP} properties of the Higgs boson interactions with gauge bosons have been performed by the ATLAS \cite{ATLAS:2016ifi,ATLAS:2017azn,ATLAS:2018hxb} and CMS \cite{CMS:2016tad,CMS:2019ekd,CMS:2019jdw} experiments, and the results show no deviations from the SM predictions.
ATLAS and CMS also finished analyses of $H t \bar{t}$  coupling, which provides an alternative and independent avenue for \textit{CP} testing in the Higgs sector because it is particularly sensitive to deviations from the SM coupling \cite{ATLAS:2020ior,CMS:2020cga};
%Add Hvv pure cp-odd.
Their results show exclusion of the pure \textit{CP}-odd structure of the top quark Yukawa ($t \bar{t} H$) coupling at $3.9\sigma$ ($3.2\sigma$) and the fractional contribution of the \textit{CP}-odd component is measured to be $f_{C P}^{H t \bar t}=0.00 \pm 0.33$. \par

However, small anomalous contributions were not excluded. At the HL-LHC~\cite{Cepeda:2019klc}, the
 $CP$-odd $VVH$ couplings are introduced and the expected results constrain the $CP$-odd parameters $\tilde{c}_{Z \gamma}$ between $-0.22$ and $0.22$ and the $\tilde{c}_{Z Z}$ between $-0.33$ and $ 0.33$ at the $68\%$ confidence level. 
\par

In terms of probing the \textit{CP}-odd Higgs couplings, a lepton collider operating as a Higgs factory has great advantages, as it is free of the QCD background and has tunable and precisely defined initial energies.
Several future lepton colliders, including the International Linear Collider (ILC) \cite{Baer:2013cma}, the $e^{+} e^{-}$ Future Circular Collider (FCC-ee) \cite{FCC}, the Compact Linear Collider (CLIC) \cite{CLIC:2016zwp}, and the Circular Electron-Positron Collider (CEPC) \cite{CEPC-SPPCStudyGroup:2015csa}, have been proposed with the capability of precise measurement of Higgs boson parameters.
For the past $CP$-odd $VVH$ couplings studies, at the ILC~\cite{ILC_CP_phase}, the $CP$-mixing  angle can reach an accuracy of $4.3^{\circ}$. The $CP$ violation parameter $\tilde{g}$ can reach the limit of
$-0.04 \sim0.01$ at the CEPC~\cite{Li:2019evl}.
As for the past $CP$ study at the CLIC~\cite{CLIC_CP}, their results show the potential of measuring $CP$ violation in the top-quark sector at future $e^{+} e^{-}$ collider~\cite{snowmass2022CP}.
\par

The CEPC will operate at a center-of-mass energy of $\sqrt{s} \sim 240\ \mathrm{GeV}$ which is close to the maximum of the Higgs boson production cross-section through the $e^{+} e^{-} \rightarrow Z H$ process. Over one million Higgs bosons with an integrated luminosity of $5.6\ \mathrm{ab}^{-1}$ will be produced.
In comparison to the LHC, the cleaner environment of the CEPC allows significantly better exclusive measurements of Higgs boson decay channels. 
So, in the future, more precise Higgs-gauge boson coupling studies can be performed, such as this Higgs \textit{CP} measurement through $e^{+} e^{-} \rightarrow Z H \rightarrow (e^{+} e^{-} /\mu^{+} \mu^{-}) H(\rightarrow b \bar{b} / c \bar{c} / g g)$ process \cite{Ioffe:1976sd}.\par

This letter is organized as follows:  In Section~\ref{sec:2}, we introduces the theory framework for the analysis of Higgs $CP$-mixing.
The MC samples and event selections are described in Section~\ref{sec:3} and Section~\ref{sec:4} respectively. Section~\ref{sec:5} describes the strategy for analysis and interprets the results of this study. The conclusions are presented in Section~\ref{sec:6}.

%-------------------------------------------------%
\section{\label{sec:2}Theory framework}
To parametrize BSM effects in a general way, we assume that the new physics sector is characterized by a scale $\Lambda$, which is significantly higher than the electroweak scale, and the SM is supplemented with 59 independent dimension-6 operators. This Lagrangian can be schematically cast as~\cite{Grzadkowski:2010es}:
%new cite here.
\begin{equation}
\label{eq.1}
\mathcal{L}_{\mathrm{eff}}=\mathcal{L}_{\mathrm{SM}}^{(4)}+\frac{1}{\Lambda^{2}} \sum_{k=1}^{59} \alpha_{k} \mathcal{O}_{k},
\end{equation}
where the $\alpha_{k}$ is the coupling of operator $\mathcal{O}_{k}$.\par
Apart from the SM tree contributions, we only consider effects of order $1 / \Lambda^{2}$ on the decay amplitude. In the broken-symmetry phase, the effective Lagrangian Eq.(\ref{eq.1}) generates the terms~\cite{Beneke:2014sba} \cite{Craig:2015wwr}:
\begin{equation}
\label{eq.2}
\begin{aligned} \mathcal{L}_{\mathrm{eff}} \supset & c_{Z Z}^{(1)} H Z_{\mu} Z^{\mu}+c_{Z Z}^{(2)} H Z_{\mu \nu} Z^{\mu \nu}+c_{Z \widetilde{Z}} H Z_{\mu \nu} \widetilde{Z}^{\mu \nu}+c_{A Z} H Z_{\mu \nu} A^{\mu \nu}+c_{A \widetilde{Z}} H Z_{\mu \nu} \widetilde{A}^{\mu \nu} \\ &+H Z_{\mu} \bar{\ell} \gamma^{\mu}\left(c_{V}+c_{A} \gamma_{5}\right) \ell+Z_{\mu} \bar{\ell} \gamma^{\mu}\left(g_{V}-g_{A} \gamma_{5}\right) \ell-g_{\mathrm{em}} Q_{\ell} A_{\mu} \bar{\ell} \gamma^{\mu} \ell,
\end{aligned}
\end{equation}
which includes the relevant tree-level SM terms. The Higgs-gauge couplings of Eq.(\ref{eq.2}) are given by 
\begin{equation}
\begin{aligned}
c_{Z Z}^{(1)} &=m_{Z}^{2}\left(\sqrt{2} G_{F}\right)^{1 / 2}\left(1+\widehat{\alpha}_{Z Z}^{(1)}\right), \\
c_{Z Z}^{(2)} &=\left(\sqrt{2} G_{F}\right)^{1 / 2} \widehat{\alpha}_{Z Z}, \\
c_{Z \widetilde{Z}} &=\left(\sqrt{2} G_{F}\right)^{1 / 2} \widehat{\alpha}_{Z \widetilde{Z}}, \\
c_{A Z} &=\left(\sqrt{2} G_{F}\right)^{1 / 2} \widehat{\alpha}_{A Z}, \\
c_{A \widetilde{Z}} &=\left(\sqrt{2} G_{F}\right)^{1 / 2} \widehat{\alpha}_{A \widetilde{Z}},
\end{aligned}
\end{equation}
where the $\hat{\alpha}_{A \tilde{Z}}$ and $\hat{\alpha}_{Z \tilde{Z}}$ are \textit{CP}-violation parameters.\par

The differential cross-section for $e^{+} e^{-} \rightarrow Z H \rightarrow l^{+} l^{-} H(\rightarrow b \bar{b} / c \bar{c} / g g)$ is given by:
\begin{equation}
\label{eq.4}
\frac{\mathrm{d} \sigma}{\mathrm{d} \cos \theta_{1} \mathrm{dcos} \theta_{2} \mathrm{~d} \psi}=\frac{1}{m_{H}^{2}} \mathcal{N}_{\sigma}\left(q^{2}\right) \mathcal{J}\left(q^{2}, \theta_{1}, \theta_{2}, \phi\right),
\end{equation}
 where the definitions of three angles are shown in Appendix~\ref{appendix1}. $\mathcal{N}_{\sigma}\left(q^{2}\right)$ is the normalization factor and it can be written in terms of the dimensionless parameters $r$ and $s$ as:
\begin{equation}
\mathcal{N}_{\sigma}\left(q^{2}\right)=\frac{1}{2^{10}(2 \pi)^{3}} \frac{1}{\sqrt{r}\ \gamma\ Z} \frac{\sqrt{\lambda(1, s, r)}}{s^{2}},
\end{equation}
the constant dimensionless parameters are given by the following:
\begin{equation}
s=\frac{q_{\mathrm{th}}^{2}}{m_{H}^{2}} \approx 3.68, r=\frac{m_{Z}^{2}}{m_{H}^{2}} \approx 0.53, \gamma Z=\frac{\Gamma_{Z}}{m_{H}} \approx 0.020 \text {, }
\end{equation}
also the $J$ can be expressed by:

\begin{equation}
\begin{aligned}
\mathcal{J}\left(q^{2}, \theta_{1}, \theta_{2}, \phi\right)=& J_{1}\left(1+\cos ^{2} \theta_{1} \cos ^{2} \theta_{2}+\cos ^{2} \theta_{1}+\cos ^{2} \theta_{2}\right) \\
&+J_{2} \sin ^{2} \theta_{1} \sin ^{2} \theta_{2}+J_{3} \cos \theta_{1} \cos \theta_{2} \\
&+\left(J_{4} \sin \theta_{1} \sin \theta_{2}+J_{5} \sin 2 \theta_{1} \sin 2 \theta_{2}\right) \sin \phi \\
&+\left(J_{6} \sin \theta_{1} \sin \theta_{2}+J_{7} \sin 2 \theta_{1} \sin 2 \theta_{2}\right) \cos \phi \\
&+J_{8} \sin ^{2} \theta_{1} \sin ^{2} \theta_{2} \sin 2 \phi+J_{9} \sin ^{2} \theta_{1} \sin ^{2} \theta_{2} \cos 2 \phi,
\end{aligned}
\end{equation}
where the explicit form of the $J_{i}$ in terms of the EFT coefficients and Standard Model parameters was computed by \cite{Beneke:2014sba} and for convenience is given in Appendix \ref{appendix}.

Among all the BSM variables, only $\hat{\alpha}_{A \tilde{Z}}$ and $\hat{\alpha}_{Z \tilde{Z}}$, which show in Eq.(\ref{A2}), contribute to the \textit{CP}-odd. So those are the \textit{CP}-violating parameters that we need to study.\par

In addition to simplify the analysis, we only constrain the $CP$-violating parameters with the assumption that all other parameters are zero.\par
Three of the $J_{i}$ functions shown in Eq.(\ref{A1}), namely $J_{4}, J_{5}, J_{8},$ are \textit{CP}-odd and vanish in the SM at tree level, whereas the remaining six functions are \textit{CP}-even.\ 
As a result, the differential cross-section can be represented as follows:
%\begin{footnotesize}
\begin{equation}
\frac{d \sigma}{d \cos \theta_{1} d \cos \theta_{2} d \phi}=N \times\left(J_{even}\left(\theta_{1}, \theta_{2}, \phi\right)+\hat{\alpha}_{A \tilde{Z}} \times J_{odd_1}\left(\theta_{1}, \theta_{2}, \phi\right)+\hat{\alpha}_{Z \tilde{Z}} \times J_{odd_2}\left(\theta_{1}, \theta_{2}, \phi\right)\right),
\end{equation}
%\end{footnotesize}
where 
%$\hat{\alpha}_{A \tilde{Z}}$ and $\hat{\alpha}_{Z \tilde{Z}}$ are \textit{CP}-violating parameters,
$J_{odd_1}$ and $J_{odd_2}$ are part of $J_{odd}$.
\par
\begin{figure}[!htbp]
\centering
\subfigure[]{
\includegraphics[width=7.5cm]{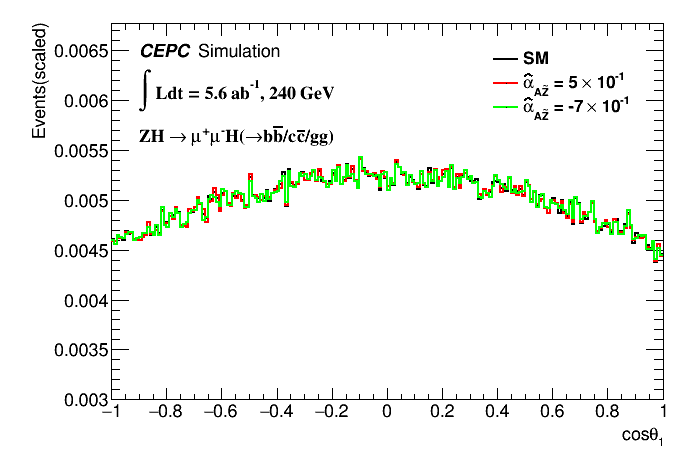}
%\caption{fig1}
}
\quad
\subfigure[]{
\includegraphics[width=7.5cm]{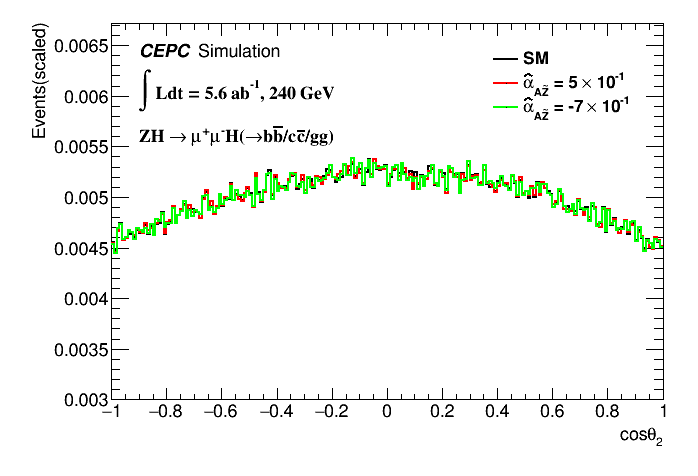}
}
\quad
\subfigure[]{
\includegraphics[width=7.5cm]{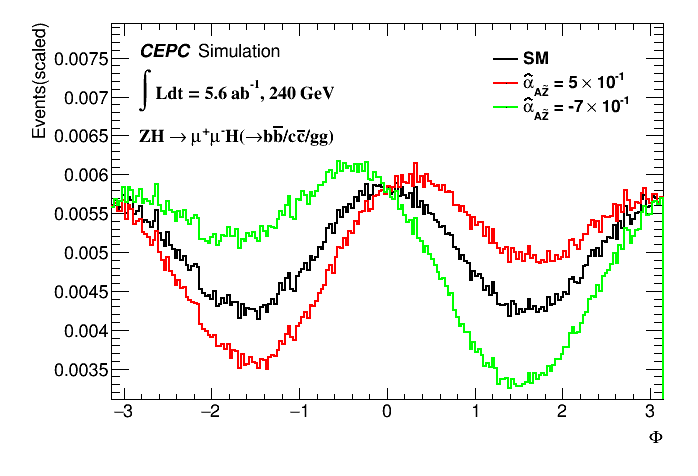}
}
\quad
\subfigure[]{
\includegraphics[width=7.5cm]{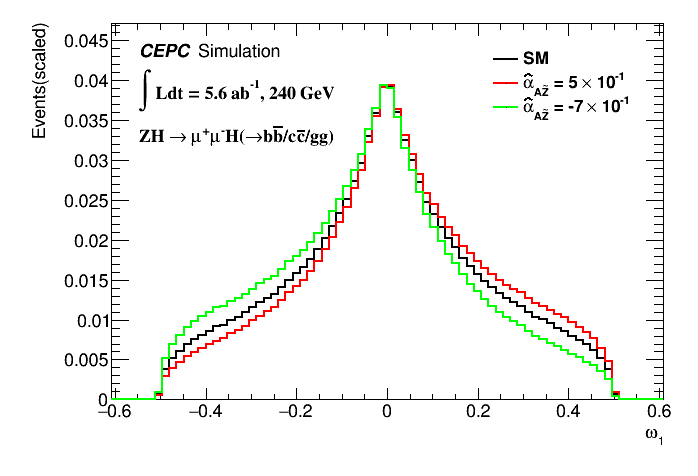}
}
%\captionsetup{font={scriptsize}}
\caption{Parametric curves with different $\hat{\alpha}_{A \tilde{Z}}$ values. The optimal variable $\omega_{1}$ is defined in the text.(The value of $\omega_{1}$ here is multiplied by 1000 for numerical convenience.)}
\label{fig:variable distribution1}
\end{figure}
\begin{figure}[!htbp]
\centering
\subfigure[]{
\includegraphics[width=7.5cm]{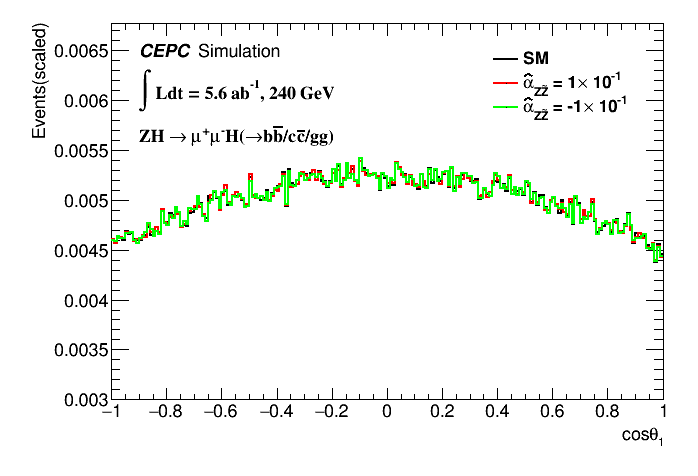}
%\caption{fig1}
}
\quad
\subfigure[]{
\includegraphics[width=7.5cm]{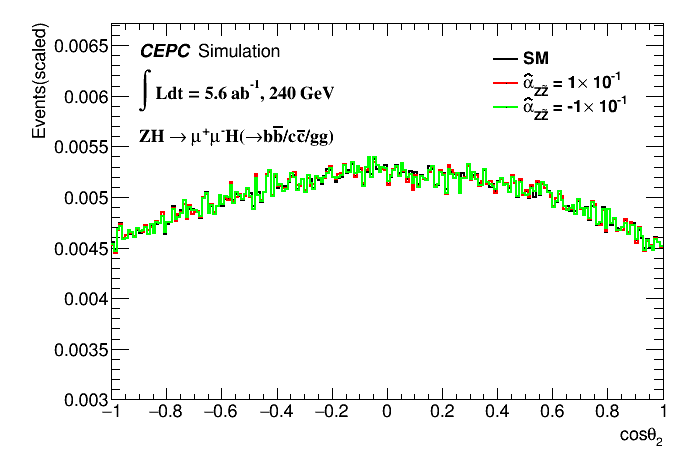}
}
\quad
\subfigure[]{
\includegraphics[width=7.5cm]{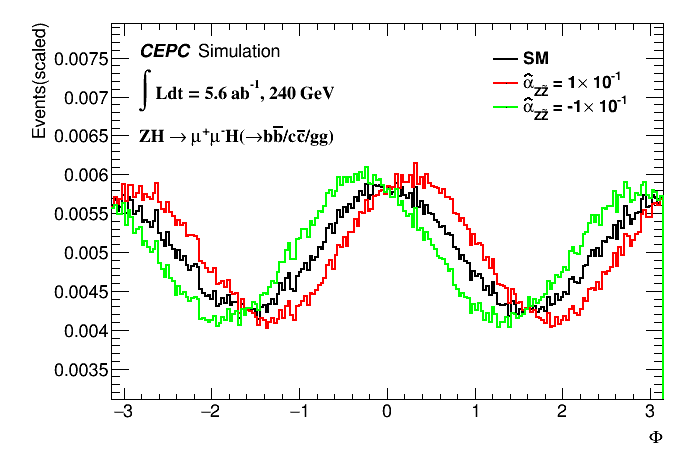}
}
\quad
\subfigure[]{
\includegraphics[width=7.5cm]{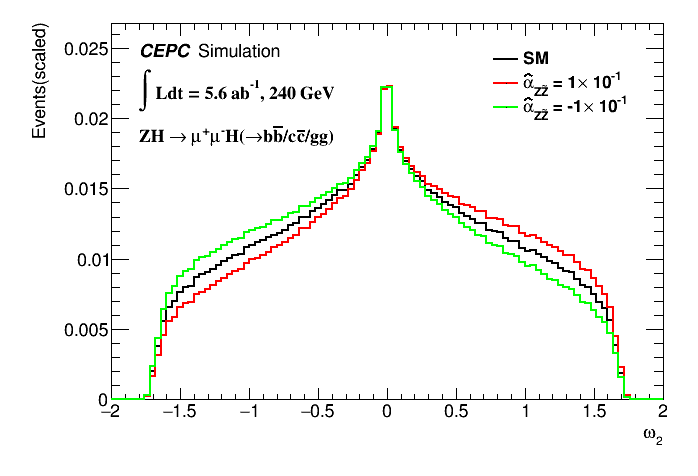}
}
%\captionsetup{font={scriptsize}}
\caption{Parametric curves with different $\hat{\alpha}_{Z \tilde{Z}}$ values. The optimal variable $\omega_{2}$ is defined in the text.(The value of $\omega_{2}$ here is multiplied by 1000 for numerical convenience.)}
\label{fig:variable distribution2}
\end{figure}

For the sake of convenience and effectiveness, two optimal variables combining the information from $\left\{\theta_{1}, \theta_{2}, \phi\right\}$ can be defined as \cite{Davier:1992nw}:
\begin{equation}
\omega_{1}=\frac{J_{odd_1}\left(\theta_{1}, \theta_{2}, \phi\right)}{J_{even}\left(\theta_{1}, \theta_{2}, \phi\right)},
\end{equation}
\begin{equation}
\omega_{2}=\frac{J_{odd_2}\left(\theta_{1}, \theta_{2}, \phi\right)}{J_{even}\left(\theta_{1}, \theta_{2}, \phi\right)},
\end{equation}
where $\omega_{1}$ and $\omega_{2}$ combine the information from 3-dimension phase space and can be used to measure $\hat{\alpha}_{A \tilde{Z}}$ and $\hat{\alpha}_{Z \tilde{Z}}$, respectively.
The parametric curves with different $\hat{\alpha}_{A \tilde{Z}}$ values are shown in Fig~\ref{fig:variable distribution1} and with different $\hat{\alpha}_{Z \tilde{Z}}$ values in Fig~\ref{fig:variable distribution2}.
\par

%-------------------------------------------------%
\section{\label{sec:3}Monte Carlo Samples}
The SM Higgs and background samples, generated with Whizard 1.95 \cite{Kilian:2007gr} and fully simulated with the CEPC baseline\cite{CEPC-SPPCStudyGroup:2015csa} detector design, are used to calculate the selection efficiencies and study background. The details of the event generation can be found at Ref.\cite{Mo:2015mza}.
Moreover, $CP$-mixing Higgs samples are generated according to Eq.(\ref{eq.1}).
The study are based on the $\sqrt{s}$ that equals to $240\ \mathrm{GeV}$.\ 
The mass of the Higgs boson is set to be $125\ \mathrm{GeV}$.\ 
All the generations are normalized to the expected yields with an integrated luminosity of $5.6\ \mathrm{ab}^{-1}$.\par

%-------------------------------------------------%
\section{\label{sec:4}Event selection}
The signal sample is the process of $e^{+} e^{-} \rightarrow Z H \rightarrow l^{+} l^{-} H(\rightarrow b \bar{b} / c \bar{c} / g g)$, containing two jets and two leptons with opposite charges in the final state. Only irreducible backgrounds are considered in the study, mainly $e^{+} e^{-} \rightarrow Z Z \rightarrow l^{+} l^{-} q \bar{q}$.\par 
Each event must contain two isolated tracks with opposite charges, reconstructed as $e^{+} e^{-}$ or $\mu^{+} \mu^{-}$ .The energy of each isolated lepton candidate must be above $20\ \mathrm{GeV}$. 
The polar angle of the muon pair system is required to be in the range of $\left|\cos \theta_{\mu^{+} \mu^{-}}\right|<0.81$. The invariant mass of the muon pair must be within the Z mass window, which is defined as from $77.5\ \mathrm{GeV}$ to $104.5\ \mathrm{GeV}$.\par
The invariant mass of the muon pair's recoil system, denoted as $M_{\text {recoil }}^{\mu^{+} \mu^{-}}$, can provide a clear signature of the $\mu {\mu} H$ events. 
The definition of $M_{\text {recoil }}^{\mu^{+} \mu^{-}}$ is: 
\begin{equation}
M_{\text {recoil }}^{\mu^{+} \mu^{-}}=\sqrt{\left(\sqrt{s}-E_{\mu^{+} \mu^{-}}\right)^{2}-p_{\mu^{+} \mu^{-}}^{2}}=\sqrt{s-2 E_{\mu^{+} \mu^{-}} \sqrt{s}+m_{\mu^{+} \mu^{-}}^{2}},
\end{equation}
in which  $\sqrt{s}=240\ \mathrm{GeV}$. 
While $E_{\mu \mu}$ and $m_{\mu \mu}$ stand for the energy and mass of the muons, respectively. A Higgs mass window is defined by requiring $M_{\text {recoil }}^{\mu^{+} \mu^{-}}$ between $124\ \mathrm{GeV}$ and $140\ \mathrm{GeV}$.
\par

The remaining particles in the event are used to reconstruct exactly two jets with a polar angle $\theta_{\text {jet }}$ in the range of $\left|\cos \theta_{\text {jet }}\right|<0.96$, using $ee$-kt algorithm. 
The invariant mass of the pair of jets is required to be between $100\ \mathrm{GeV}$ and $150\ \mathrm{GeV}$ to reject the background.\par

Compared to the analysis of the $Z H \rightarrow \mu^{+} \mu^{-} H$ decay, the analysis of the $Z H\rightarrow e^{+} e^{-} H$ decay suffers from large background. A cut based event selection is performed for the $ Z H \rightarrow e^{+} e^{-} H$ process. The electron-positron pair is required to have its invariant mass in the range of $85-95\ \mathrm{GeV}$ and the polar angle of each electron is required to satisfy  $\left|\cos \phi_{e}\right|<0.95$. The other selection criteria are same as $\mu^{+} \mu^{-} H$ analysis. It should be noticed that the effect of $Z$-fusion in $e^+e^-H$ process is neglected in this study since its cross section is rather small. \par

The expected signal and background yields during the event selections are summarized in Tab~\ref{tab1} for $\mu^{+} \mu^{-} H$ and $e^{+} e^{-} H$ analysis, respectively.

\begin{table}[!htbp]
%\captionsetup{font={scriptsize}}
\caption{\label{tab1} Event yields of cut flow. Signal events are $Z H \rightarrow l^{+} l^{-} H$, $H\rightarrow b \bar{b} / c \bar{c} / g g$ combined. Background is the $e^{+} e^{-} /\mu^{+} \mu^{-}+$jet pair process.}
\begin{tabular}{p{0.25 \textwidth}<{}p{0.25 \textwidth}<{\centering}p{0.25 \textwidth}<{\centering}}
\hline
\multicolumn{3}{c}{$Z H \rightarrow \mu^{+} \mu^{-} H(\rightarrow b \bar{b} / c \bar{c} / g g)$ channel}     \\ \hline
                    &Signal   &Background   \\ \hline
Original            & $2.62 \times 10^{4}$ &$1.25 \times 10^{6}$ \\ \hline
Leptonn pair selection & $1.59 \times 10^{4}$ &  $9.91 \times 10^{3}$ \\ \hline
All selection       & $1.48 \times 10^{4}$ & $5.60 \times 10^{3}$ \\ \hline
\multicolumn{3}{c}{$Z H \rightarrow e^{+} e^{-} H(\rightarrow b \bar{b} / c \bar{c} / g g)$ channel}     \\ \hline
                    &Signal   &Background   \\ \hline
Original            & $2.72 \times 10^{4}$ &$1.77 \times 10^{6}$ \\ \hline
Lepton pair selection & $8.76 \times 10^{3}$ &  $8.77 \times 10^{4}$ \\ \hline
All selection       & $7.15 \times 10^{3}$ & $4.59 \times 10^{3}$ \\ \hline
\end{tabular}
\end{table}

%-------------------------------------------------%
\section{\label{sec:5}Fitting strategy and Result}
After the event selections, the correlations among ($\theta_{1}, \theta_{2}, \phi$) and the variables for selection, such as $\cos \theta_{l^{+} l^{-}}, \operatorname{Mass}_{l^{+} l^{-}}, M_{\text {recoil }}^{l^{+} l^{-}}, \cos \theta_{\mathrm{jet}} ,\operatorname{Mass}_{j j}$, are carefully investigated and the impacts on $CP$ study are negligible. 

%-------------------------------------------------%
\subsection{\label{sec:5_1}$\mu^{+} \mu^{-} H$ results obtained by $\omega$-fitting}
The \textit{CP}-violating parameters $\hat{\alpha}_{A \tilde{Z}}$ and $\hat{\alpha}_{Z \tilde{Z}}$ can be measured through the optimal variable $\omega_{1}$ and $\omega_{2}$. The estimation of $\hat{\alpha}_{A \tilde{Z}}$ and $\hat{\alpha}_{Z \tilde{Z}}$ uses a maximum-likelihood fit which could be constructed as:
\begin{equation}
\mathcal{L}(\vec{x} \mid \vec{\alpha})=\prod_{\text {data }} f\left(x_{i} \mid \vec{\alpha}\right),
\end{equation} 
where $\vec{\alpha}$ are \textit{CP}-violating parameters ($\hat{\alpha}_{A \tilde{Z}}$ and $\hat{\alpha}_{Z \tilde{Z}}$) to be estimated, and $x$ is the dataset.
For each $\vec{\alpha}$ hypothesis, the profile of a negative log-likelihood (NLL) is calculated. 
The best-estimated $\vec{\alpha}$, as well as its central confidence interval at a $68 \% \ (95 \%)$ confidence level (CL), can be determined at $\Delta N L L=N L L-N L L_{\min }= 0.5 \  (1.96)$.
\par

The main sensitive variable for the \textit{CP} test in this analysis is the optimal variable $\omega$ (stand for $\omega_{1}$ and $\omega_{2}$), which combines all 3 kinematic variables of $\theta_{1}, \theta_{2},$ and $\phi$, and the function used to determine the $\vec{\alpha}$-value can be defined as:
\begin{equation}
\label{function:f_p}
f^{\vec{\alpha}}(\omega)=N_{\mathrm{sig}} * f_{\mathrm{sig}}^{\vec{\alpha}}(\omega)+N_{b k g} * f_{b k g}^{\vec{\alpha}}(\omega) ,
\end{equation}
where $f_{\text {sig }}^{\vec{\alpha}}(\omega)$ and $f_{\text {bkg }}^{\vec{\alpha}}(\omega)$ are probability density functions (PDFs) of the signal and background, and $N_{\text {sig}}$ and $N_{\text {bkg}}$ are the yields of them, respectively.
 
\begin{figure}[!htbp]
\centering
\label{Fig: fit_plot}
\subfigure[]{
\includegraphics[width=7.5cm]{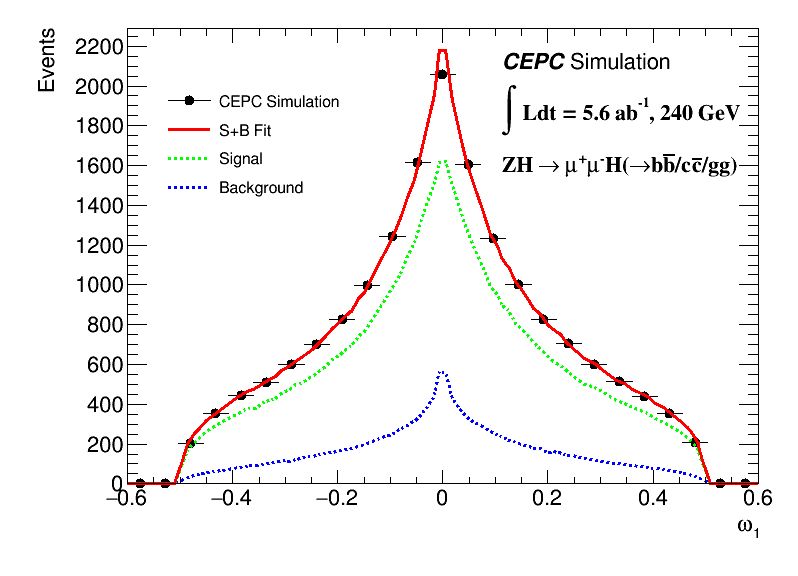}
}
\quad
\subfigure[]{
\includegraphics[width=7.5cm]{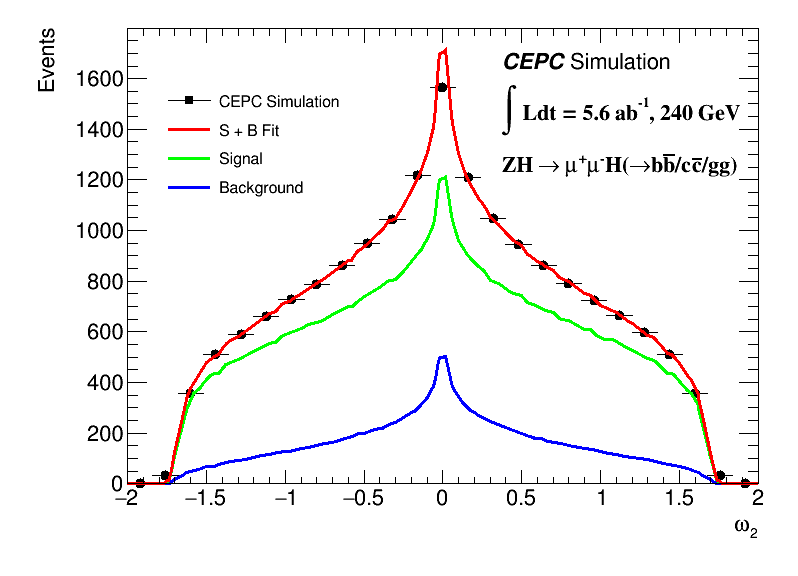}
}
\quad
\subfigure[]{
\includegraphics[width=7.5cm]{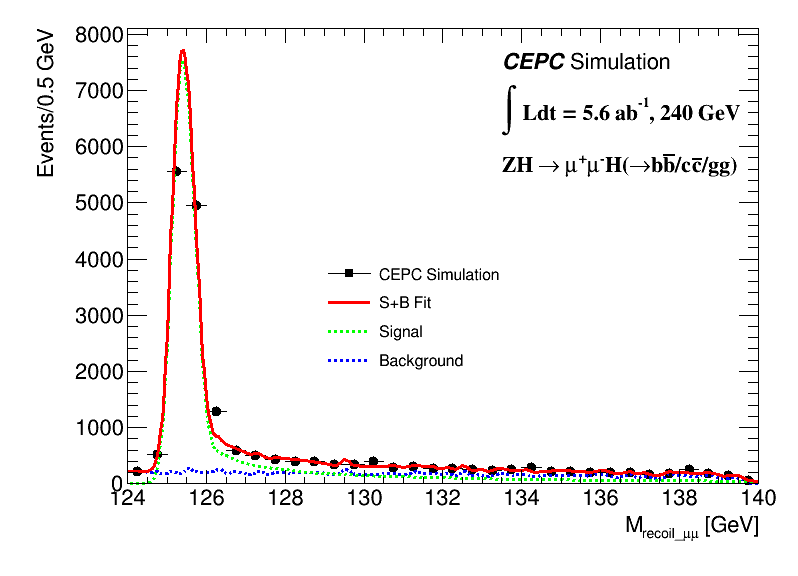}
}
%\captionsetup{font={scriptsize}}
\caption{(a) Fit result projected on $\omega_{1}$ distribution in $\mu^+\mu^- H$ channel, (b) Fit result projected on $\omega_{2}$ distribution in $\mu^+\mu^- H$ channel, (c) Fit result projected on recoil mass distribution in $\mu^+\mu^- H$ channel. }
\label{fig2}
\end{figure}

For the modeling of signal, PDFs are generated according to different $\vec{\alpha}$ hypothesises, and background PDF is fixed to the MC simulation. 
The $M_{\text {recoil }}^{\mu^{+} \mu^{-}}$ distribution is essential to discriminate signal over background. The background, dominated by the $e^{+} e^{-} \rightarrow Z Z \rightarrow \mu^+\mu^- q \bar{q}$, is modeled by a second-order polynomial, while the signal is modeled by the Crystal Ball function. Fig~\ref{fig2} shows the fit result of $\omega$ and $M_{\text {recoil }}^{\mu^{+} \mu^{-}}$.
\par

\begin{figure}[!htbp]
\centering
\subfigure[]{
\includegraphics[width=7.5cm]{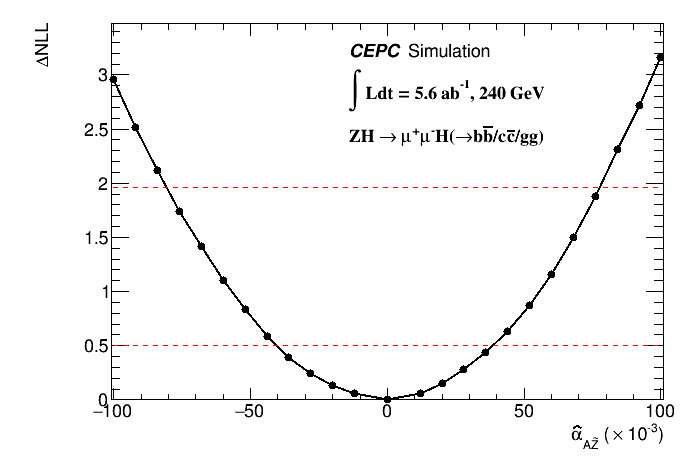}
}
\quad
\subfigure[]{
\includegraphics[width=7.5cm]{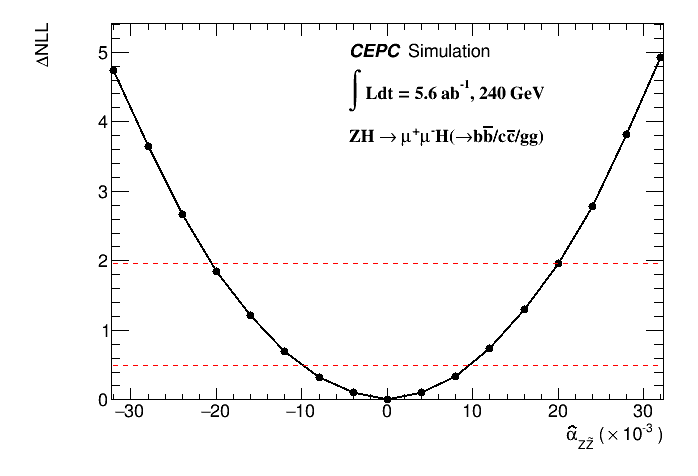}
}
\caption{(a) $\Delta N L L$ curve (fit to $\hat{\alpha}_{A \tilde{Z}}$ with $\widehat{\alpha}_{Z \tilde{Z}}=0$), (b) $\Delta N L L$ curve (fit to $\hat{\alpha}_{Z \tilde{Z}}$ with $\widehat{\alpha}_{A \tilde{Z}}=0$). }
\label{fig4}
\end{figure}

The individual fitting with each single CPV parameter has been performed by assuming $\widehat{\alpha}_{A \tilde{Z}}$ is a free parameter with $\widehat{\alpha}_{Z \tilde{Z}}=0$ and vice versa. The expected and observed $\Delta NLL$ curves are shown in Fig~\ref{fig4} as a quadratic function of $\vec{\alpha}$.
It corresponds to the SM prediction that $\vec{\alpha}$ equals to zero,
and the results of $\widehat{\alpha}_{Z \tilde{Z}}$ and $\hat{\alpha}_{A \tilde{Z}}$ with confidence interval at  $68 \%$ ($95 \%$) 
represent the sensitivity to a $CP$-odd Higgs, which are shown in Tab~\ref{tab:phi_c}. \par

%\clearpage

%-------------------------------------------------%
\subsection{\label{sec:5_2}$\mu^{+} \mu^{-} H$ results obtained by $\phi$-fitting}
Because the $\phi$ contains the most information among the three kinematic variables, it is straightforward and feasible to fit to $\phi$. \par 
 \begin{figure}[!htbp]
 \centering
 \includegraphics[width=8cm]{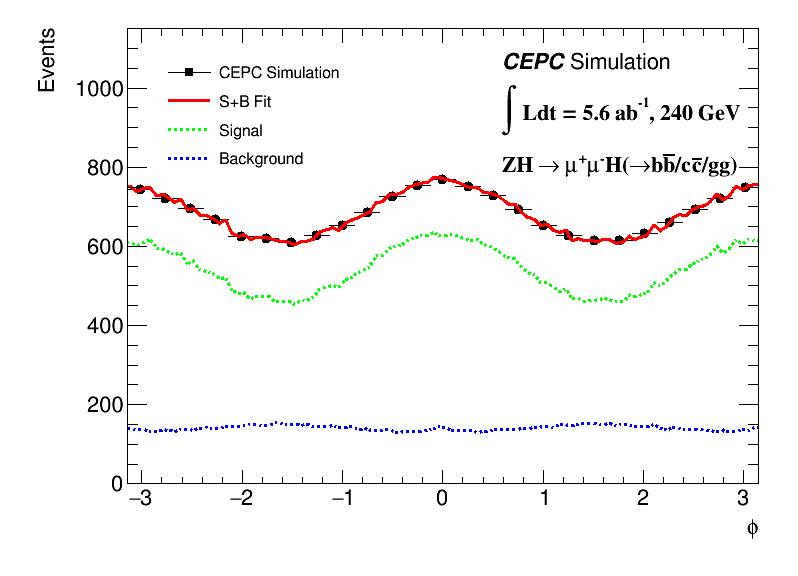}
 %\captionsetup{font={scriptsize}}
 \caption{\label{fig:Fit_phi}Fit result projected on $\phi$ distribution in $\mu^+\mu^- H$ channel.}
 \end{figure}
 
Similar to the Eq.(\ref{function:f_p}), the function used to obtain the $\vec{\alpha}$ is as following:
\begin{equation}
\label{function:f_p_phi}
f^{\vec{\alpha}}(\phi)=N_{\text {sig }} * f_{\text {sig }}^{\vec{\alpha}}(\phi)+N_{b k g} * f_{b k g}^{\vec{\alpha}}(\phi),
\end{equation}
where the definitions of $f_{\text {sig}}^{\vec{\alpha}}(\phi)$, $f_{\text {bkg }}^{\vec{\alpha}}(\phi)$, $N_{\text {sig}}$ and $N_{\text {bkg}}$ are the same as those used to fit to $\omega$ above.\par

\begin{figure}[!htbp]
\centering
\subfigure[]{
\includegraphics[width=7.5cm]{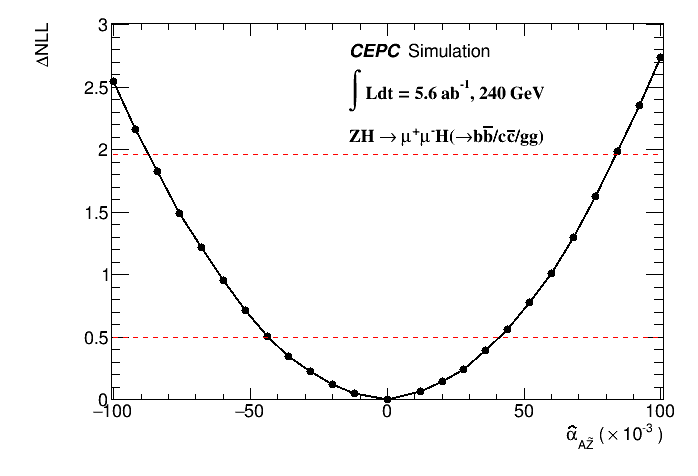}
}
\quad
\subfigure[]{
\includegraphics[width=7.5cm]{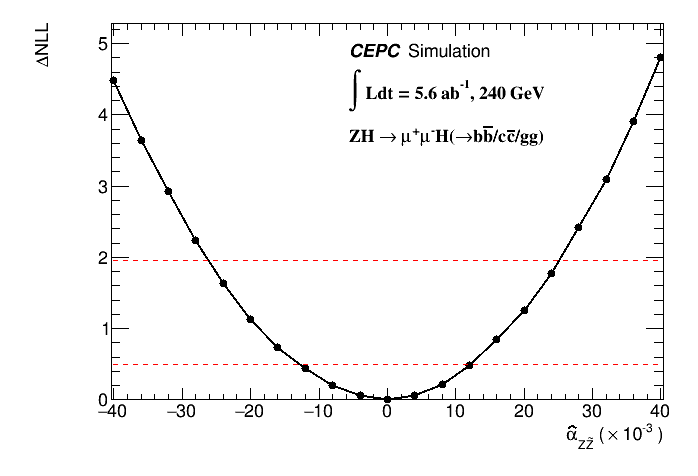}
}
\caption{(a) $\Delta N L L$ curve (fit to $\hat{\alpha}_{A \tilde{Z}}$ with $\widehat{\alpha}_{Z \tilde{Z}}=0$), (b) $\Delta N L L$ curve (fit to $\hat{\alpha}_{Z \tilde{Z}}$ with $\widehat{\alpha}_{A \tilde{Z}}=0$). }
\label{fig5}
\end{figure}

After fitting with a single $CPV$ parameter, the fit results are shown in Fig~\ref{fig:Fit_phi}, the $\Delta N L L$ curves in Fig~\ref{fig5}.

The comparison with the results of $\omega$-fitting is shown in Tab~\ref{tab:phi_c}. It can be seen that the results of the $\phi$-fitting is slightly worse than those of the $\omega$-fitting as expected, since fewer kinematic variables, i.e, less information, are used.

\begin{table}[!htbp]
\caption{\label{tab:phi_c} Summary of 1$\sigma$ and 2$\sigma$ bounds on $\hat{\alpha}_{A \tilde{Z}}$ and $\hat{\alpha}_{Z \tilde{Z}}$ from various analyses by fitting to $\phi$ and fitting to $\omega$ which is shown in Sec~\ref{sec:5_1}.}

\begin{tabular}{p{0.2 \textwidth}<{}|p{0.2 \textwidth}<{\centering}p{0.2 \textwidth}<{\centering}}
\hline
& $\hat{\alpha}_{A \tilde{Z}} (\times 10^{-2})$ & $\hat{\alpha}_{Z \tilde{Z}} (\times 10^{-2})$ \\ \hline
&\multicolumn{2}{c}{$\omega$-fitting} \\ \hline
$68\%$ CL(1$\sigma$)            & $\left[-4.16 , 3.88 \right]$ & $\left[-1.06 , 1.00 \right]$\textbf{} \\ \hline
$95\%$ CL(2$\sigma$) & $\left[-8.10 , 7.82 \right]$ & $\left[-2.06 , 2.01 \right]$ \\ \hline
&\multicolumn{2}{c}{$\phi$-fitting} \\ \hline
$68\%$ CL(1$\sigma$) & $\left[-4.42 , 4.21 \right]$ & $\left[-1.35 , 1.24 \right]$\textbf{} \\ \hline
$95\%$ CL(2$\sigma$) & $\left[-8.66 , 8.45 \right]$ & $\left[-2.62 , 2.51 \right]$ \\ \hline

\end{tabular}
\end{table}

\subsection{\label{sec:5_3}Combined results obtained by $\omega$-fitting}
By using the same process shown in \ref{sec:5_1}, the $e^{+} e^{-} H$ results with $\omega$-fitting can be easy obtained.
Neglecting the migration between $e^{+} e^{-} H$ and $\mu^{+} \mu^{-} H$, the combined likelihood is equal to the multiplication of the likelihood of $e^{+} e^{-} H$ and $\mu^{+} \mu^{-} H$ channels.
\par
To compare with HL-LHC \cite{Cepeda:2019klc}, it is necessary to do some conversion. The CPV Lagrangian which HL-LHC used is given by,
\begin{footnotesize}
\begin{equation}
\label{eq.16}
\mathcal{L}_{\mathrm{CPV}}=\frac{H}{v}\left[\tilde{c}_{\gamma \gamma} \frac{e^{2}}{4} A_{\mu \nu} \tilde{A}^{\mu \nu}+\tilde{c}_{Z \gamma} \frac{e \sqrt{g_{1}^{2}+g_{2}^{2}}}{2} Z_{\mu \nu} \tilde{A}^{\mu \nu}+\tilde{c}_{Z Z} \frac{g_{1}^{2}+g_{2}^{2}}{4} Z_{\mu \nu} \tilde{Z}^{\mu \nu}+\tilde{c}_{W W} \frac{g_{2}^{2}}{2} W_{\mu \nu}^{+} \tilde{W}^{-\mu \nu}\right],
\end{equation}
\end{footnotesize}
where $g_{1}$ and $g_{2}$ are the $U(1)_{Y}$ and $S U(2)_{L}$ gauge coupling constants. 
\par
Comparing Eq.(\ref{eq.16}) and Eq.(\ref{eq.2}), the connections of the \textit{CP}-odd related parameters between the two equations are obvious,
\begin{equation}
\begin{aligned}
\label{eq.17}
&\left(\sqrt{2} G_{F}\right)^{1 / 2} \hat{\alpha}_{Z Z} H Z_{\mu v} \tilde{Z}^{\mu v}=\frac{H}{v} \tilde{c}_{Z Z} \frac{g_{1}^{2}+g_{2}^{2}}{4} Z_{\mu v} \tilde{Z}^{\mu v}, \\
&\left(\sqrt{2} G_{F}\right)^{1 / 2} \hat{\alpha}_{A \tilde{Z}} H Z_{\mu \nu} \tilde{A}^{\mu \nu}=\frac{H}{v} \tilde{c}_{Z \gamma} \frac{\mathrm{e} \sqrt{g_{1}^{2}+g_{2}^{2}}}{2} Z_{\mu \nu} \tilde{A}^{\mu v},
\end{aligned}
\end{equation}
where $g_{1}$ and $g_{2}$ equal to 0.358 and 0.648, respectively, and $e$ is the EM coupling which equal to 0.313. \par

Converted with the Eq.(\ref{eq.17}), all the results are summarized in Tab \ref{tab:compare}. The $1\sigma$ bounds on $\tilde{c}_{Z \gamma}$ and $\tilde{c}_{Z Z}$ are $\left[-0.30, 0.27\right]$ and $\left[-0.06, 0.06\right]$ respectively. These results are significantly better than HL-LHC on the $\tilde{c}_{Z Z}$ and comparable on the $\tilde{c}_{Z \gamma}$.

%The combined numerical results of each channel are shown in Tab~\ref{tab:combine}.
\par
According to the latest note~\cite{snowmass2021}, CEPC Higgs operation can be upgraded to $20\ \mbox{ab}^{-1}$. With this increasing luminosity, the updated numerical results are shown in Tab~\ref{tab:combine}. This result is about two times better than that of $5.6\ \mbox{ab}^{-1}$.
\par
In order to compare with the results using amplitude. We can use the equations in ~\cite{Andrei_g4withczz}. 
%After limiting the couplings to real-valued numbers, the generality of amplitude paramterization can uniquely represent each EFT coefficient in Eq.(\ref{eq.16}):

\begin{equation}
\begin{aligned}
\label{eq.18}
\tilde{c}_{z z}=-\frac{2 s_{w}^{2} c_{w}^{2}}{e^{2}} g_{4}^{z z}, \\
\tilde{c}_{z \gamma}=-\frac{2 s_{w} c_{w}}{e^{2}} g_{4}^{z \gamma},
\end{aligned}
\end{equation}
so we can easily reach the limit of $g_{4}^{z z}$ to  $ -0.015 \sim 0.015$ at  $68\%$ confidence level with $5.6\ \mbox{ab}^{-1}$.

\begin{table}[!htbp]
\caption{\label{tab:compare} Summary of 1$\sigma$ bounds on $\tilde{c}_{Z \gamma}$ and $\tilde{c}_{Z Z}$ from various analyses considered in our study and HL-LHC analysis.}
\begin{tabular}{p{0.18 \textwidth}<{}|p{0.18 \textwidth}<{\centering}p{0.18 \textwidth}<{\centering}p{0.18 \textwidth}<{\centering}}
\hline
\hline
Collider    
        & $pp$ & $e^{+}e^{-}$  &  $e^{+}e^{-}$\\
$E \left(\mathrm{GeV}\right)$ 
        & 14000 & 240 & 240 \\
$\mathcal{L} \left(\mathrm{fb}^{-1}\right)$   
        & 3000 & 5600 & 20000 \\
\hline
$\tilde{c}_{Z \gamma} \left(1\sigma \right)$   & $\left[ -0.22, 0.22 \right]$ & $\left[ -0.30, 0.27 \right]$ & $\left[ -0.16, 0.14 \right]$ \\
\hline
$\tilde{c}_{Z Z} \left(1\sigma \right)$        & $\left[ -0.33, 0.33 \right]$ & $\left[ -0.06, 0.06 \right]$ & $\left[ -0.03, 0.03 \right]$  \\ 
\hline
\hline

\end{tabular}
\end{table}

\begin{table}[!htbp]
\caption{\label{tab:combine} Summary of 1$\sigma$ and 2$\sigma$ bounds on  $CP$-violating parameters $\tilde{c}_{Z \gamma}$, $\tilde{c}_{Z Z}$ from various analyses considered in our study with different channels by fitting to $\omega$ with $5.6\ \mbox{ab}^{-1}$ and $20\ \mbox{ab}^{-1}$.}

\begin{tabular}{p{0.15 \textwidth}<{}|p{7em}<{\centering}p{7em}<{\centering}p{7em}<{\centering}p{7em}<{\centering}}
\hline
&\multicolumn{2}{c}{$5.6\ \mbox{ab}^{-1}$}  &\multicolumn{2}{c}{$20\ \mbox{ab}^{-1}$}\\ 
\hline
& $\tilde{c}_{Z \gamma}$ & $\tilde{c}_{Z Z}$ & $\tilde{c}_{Z \gamma}$ & $\tilde{c}_{Z Z}$  \\ 
\hline
&\multicolumn{4}{c}{$\mu^{+} \mu^{-} H$ channel} \\ 
\hline
$68\%$ CL(1$\sigma$)    & $\left[-0.36, 0.33\right]$ & $\left[-0.08, 0.07\right]$       & $\left[-0.19, 0.17\right]$ & $\left[-0.04, 0.04\right]$\textbf{} \\ 
\hline
$95\%$ CL(2$\sigma$)    & $\left[-0.70, 0.67\right]$ & $\left[-0.15, 0.15\right]$       & $\left[-0.37, 0.35\right]$ & $\left[-0.08, 0.08\right]$ \\ 
\hline
&\multicolumn{4}{c}{$e^{+} e^{-} H$ channel} \\ 
\hline
$68\%$ CL(1$\sigma$)    & $\left[-0.51, 0.47\right]$ & $\left[-0.11, 0.11\right]$       & $\left[-0.28, 0.24\right]$ & $\left[-0.06, 0.06\right]$\textbf{} \\ 
\hline
$95\%$ CL(2$\sigma$)    & $\left[-1.00, 0.95\right]$ & $\left[-0.21, 0.21\right]$       & $\left[-0.53, 0.49\right]$ & $\left[-0.11, 0.11\right]$ \\ 
\hline
&\multicolumn{4}{c}{Combined results} \\ 
\hline
$68\%$ CL(1$\sigma$)    & $\left[-0.30, 0.27\right]$ & $\left[-0.06, 0.06\right]$       & $\left[-0.16, 0.14\right]$ & $\left[-0.03, 0.03\right]$\textbf{} \\ 
\hline
$95\%$ CL(2$\sigma$)    & $\left[-0.58, 0.55\right]$ & $\left[-0.12, 0.12\right]$       & $\left[-0.31, 0.28\right]$ & $\left[-0.06, 0.06\right]$ \\ 
\hline

\end{tabular}
\end{table}

%\clearpage
\section{\label{sec:6}Conclusion}
In summary, the Higgs \textit{CP} is studied by analyzing the $e^{+} e^{-} \rightarrow Z H \rightarrow (\mu^{+} \mu^{-}/e^{+} e^{-}) H(\rightarrow b \bar{b} / c \bar{c} / g g)$ process in a $5.6\ \mathrm{ab}^{-1}$  $e^{+} e^{-}$ collision sample with $\sqrt{s} = $   $240\ \mathrm{GeV}$ at the CEPC. The simplest \textit{CP} mixing model and two optimal variables combining three related kinematic variables are used in this analysis and show very promising sensitivity. The optimal variables improve results compared to just using the angular distribution between the H and Z decay-planes.
The $CP$-violating parameter $\tilde{c}_{Z \gamma}$ is determined to be greater (less) than $0.30$ ($-0 .27$) and $\tilde{c}_{Z Z}$ greater (less) than $0.06$ ($-0.06$) at $95\%$ confidence level.\par
Considering possibly the increasing luminosity such as $20\ \mathrm{ab}^{-1}$, the sensitivities to new physics could be further improved, The $CP$-violating parameter $\tilde{c}_{Z \gamma}$ is determined to be greater (less) than $0.16$ ($-0 .14$) and $\tilde{c}_{Z Z}$ greater (less) than $0.03$ ($-0.03$) at $68\%$ confidence level. In the $\tilde{c}_{Z Z}$ part, there is an order of magnitude improvement over the HL-LHC.

\acknowledgements{This study is supported by National Natural Science Foundation of China (NSFC) under grant No.~12035008 and No.~12075271.}

\clearpage
 
\appendix

\section{\label{appendix1}}
\begin{figure}[!htbp]
 \centering
 \includegraphics[width=8cm]{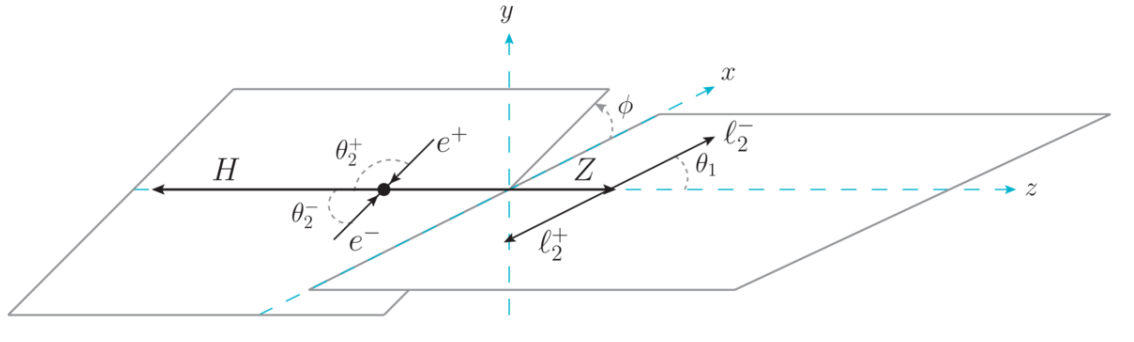}
 %\captionsetup{font={scriptsize}}
 \caption{\label{fig:xsec}Kinematics for the scattering $e^{+} e^{-} \rightarrow Z H \rightarrow l^{+} l^{-}$~\cite{Beneke:2014sba}.}
 \end{figure}

Here we describe the angle conventions used in our results.
Using the conventions for the axes giving in Fig~\ref{fig:xsec}.
We choose the direction $z$ direction to be defined by the momentum of the on-shell $Z$ boson in the $e^{+}e^{-}$ state rest frame. The $\theta_{1}$ is the angle between the momentum of $\ell^{-}$, and the $z$ axis.
The angle  $\theta_{2}^{-}$ is the angle between the direction pf flight of the $e^{-}$ and the $z$ axis in the $e^{+}e^{-}$ rest frame. To best exploit the crossing symmetry of the two processes, one should describe the reaction using the angle $\theta_{2}^{+}$ measured from the $z$ axis to the direction of flight of the $e^{+}$. Our eq.~\ref{eq.4} are therefore written in terms of the angle:
\begin{equation}
\theta_{2}^{+} \equiv \theta_{2}=\pi-\theta_{2}^{-}.
\end{equation}
Also $\phi$ is the angle between the normal of the planes defined by the $z$ direction and the momenta of $\ell^{-}$ and $e^{-}$. It is measured positively from the $\ell^{+}\ell^{-}$ plane to the $e^{+}e^{-}$ plane.

\section{\label{appendix}}
For completeness, here we list the various $J_{i}$ coefficients computed first in \cite{Beneke:2014sba}. These coefficients are conveniently expressed in terms of componets of the matrix element as 
\begin{equation}
\begin{aligned}
\label{A1}
&J_{1}=2\ r\ s\left(g_{A}^{2}+g_{V}^{2}\right)\left(\left|H_{1, V}\right|^{2}+\left|H_{1, A}\right|^{2}\right), \\
&J_{2}=\kappa\left(g_{A}^{2}+g_{V}^{2}\right)\left[\kappa\left(\left|H_{1, V}\right|^{2}+\left|H_{1, A}\right|^{2}\right)+\lambda \operatorname{Re}\left(H_{1, V}\  H_{2, V}^{*}+H_{1, A}\ H_{2, A}^{*}\right)\right], \\
&J_{3}=32 r\ s\ g_{A}\ g_{V} \operatorname{Re}\left(H_{1, V}\ H_{1, A}^{*}\right), \\
&J_{4}=4 \kappa \sqrt{r\ s\ \lambda}\ g_{A}\ g_{V} \operatorname{Re}\left(H_{1, V}\ H_{3, A}^{*}+H_{1, A}\ H_{3, V}^{*}\right), \\
&J_{5}=\frac{1}{2} \kappa \sqrt{r\ s\ \lambda}\left(g_{A}^{2}+g_{V}^{2}\right) \operatorname{Re}\left(H_{1, V}\ H_{3, V}^{*}+H_{1, A}\ H_{3, A}^{*}\right), \\
&J_{6}=4 \sqrt{r\ s}\ g_{A}\ g_{V}\left[4 \kappa \operatorname{Re}\left(H_{1, V}\ H_{1, A}^{*}\right)+\lambda \operatorname{Re}\left(H_{1, V}\ H_{2, A}^{*}+H_{1, A}\ H_{2, V}^{*}\right)\right], \\
&J_{7}=\frac{1}{2} \sqrt{r\ s}\left(g_{A}^{2}+g_{V}^{2}\right)\left[2 \kappa\left(\left|H_{1, V}\right|^{2}+\left|H_{1, A}\right|^{2}\right)+\lambda \operatorname{Re}\left(H_{1, V}\ H_{2, V}^{*}+H_{1, A}\ H_{2, A}^{*}\right)\right], \\
&J_{8}=2\ r\ s \sqrt{\lambda}\left(g_{A}^{2}+g_{V}^{2}\right) \operatorname{Re}\left(H_{1, V}\ H_{3, V}^{*}+H_{1, A}\ H_{3, A}^{*}\right), \\
&J_{9}=2\ r\ s\left(g_{A}^{2}+g_{V}^{2}\right)\left(\left|H_{1, V}\right|^{2}+\left|H_{1, A}\right|^{2}\right) .
\end{aligned}
\end{equation}

The expressions for $H_{i, V / A}$  at $\mathcal{O}\left(1 / \Lambda^{2}\right)$ are: 
\begin{equation}
\begin{aligned}
\label{A2}
&H_{1, V}=-\frac{2 m_{H}\left(\sqrt{2} G_{F}\right)^{1 / 2} r}{r-s} g_{V}\left(1+\hat{\alpha}_{1}^{\mathrm{eff}}-\frac{\kappa}{r} \hat{\alpha}_{Z Z}-\frac{\kappa}{2 r} \frac{Q_{l}\ g_{e m} (r-s)}{s\ g_{V}} \hat{\alpha}_{A Z}\right), \\
&H_{1, A}=\frac{2 m_{H}\left(\sqrt{2} G_{F}\right)^{1 / 2} r}{r-s} g_{A}\left(1+\hat{\alpha}_{2}^{\text {eff }}-\frac{\kappa}{r} \hat{\alpha}_{Z Z}\right), \\
&H_{2, V}=-\frac{2 m_{H}\left(\sqrt{2} G_{F}\right)^{1 / 2} }{r-s} g_{V}\left(2 \hat{\alpha}_{Z Z}-\frac{Q_{l}\ g_{c m}\ (r-s)}{s\ g_{V}} \hat{\alpha}_{A Z}\right), \\
&H_{2, A}=\frac{4 m_{H}\left(\sqrt{2} G_{F}\right)^{1 / 2} }{r-s} g_{A}\ \hat{\alpha}_{Z Z}\ , \\
&H_{3, V}=-\frac{2 m_{H}\left(\sqrt{2} G_{F}\right)^{1 / 2} }{r-s} g_{V}\left(2 \hat{\alpha}_{Z \tilde{Z}}+\frac{Q_{l}\  g_{e m}(r-s)}{s\ g_{V}} \hat{\alpha}_{A \tilde{Z}}\right), \\
&H_{3, A}=\frac{4 m_{H}\left(\sqrt{2} G_{F}\right)^{1 / 2} }{r-s} g_{A}\ \hat{\alpha}_{Z \tilde{Z}}\ .
\end{aligned}
\end{equation}
\par
%\begin{acknowledgments}
%\end{acknowledgments}

%\input{cp-note.bbl}
\bibliographystyle{unsrt}
\bibliography{cp-note}

\end{document}